\documentclass[prc,preprint]{revtex4}

\usepackage{graphicx}

\begin{document}

\title{\bf The effect of dynamical quark mass in the calculation of strange quark star structure}

\author{{ Gholam Hossein Bordbar$^{1,2}$
\footnote{Corresponding author. E-mail:
bordbar@physics.susc.ac.ir}} and { Babak Ziaei$^{1}$}}
\affiliation{Department of Physics,
Shiraz University, Shiraz 71454, Iran\\
and\\ Research Institute for Astronomy and Astrophysics of Maragha,
P.O. Box 55134-441, Maragha 55177-36698, Iran}
\begin{abstract}
 We have discussed dynamical behavior of strange quark matter
components, in particular the effects of density dependent quark
mass on the equation of state of strange quark matter. Dynamical
masses of quarks have been computed within Nambu-Jona-Lasinio
(NJL) model, then we have done the strange quark matter
calculations employing the MIT bag model with these dynamical
masses. For the sake of comparing dynamical mass interaction with
QCD quark-quark interaction, we have considered the
one-gluon-exchange term as the effective interaction between
quarks for MIT bag model. Our dynamical approach illustrates an
improvement for the obtained values of equation of state. We have also
investigated the structure of strange quark star using
Tolman-Oppenheimer-Volkoff (TOV) equations for all applied models.
Our results show that the dynamical mass interaction leads to
lower values for the gravitational mass.
\end{abstract}
\maketitle

\section{Introduction}
Strange quark stars (SQS) are the most compact objects with a
surface density  $\rho\sim 10^{15} \frac{gr}{cm^3}$, which is
about fourteen orders of magnitude greater than the surface
density of neutron stars, while their central density could be up
to five times higher than that (Haensel et al.~\cite{Haensel2007};
Glendenning~\cite{Glendenning2000}; Weber~\cite{Weber1999}). It
was first Itoh (1970) that, even before QCD full development,
proposed SQSs which is made of strange quark matter (SQM). Later,
Bodmer (1971) discussed the fate of an astronomical object
collapsing to such a state of matter.

The quark deconfinement hypothesis is one of the exciting steps
in investigation for the building blocks of matter. Soon after
predictions of quarks in theories and successful
laboratory observations, many hadronic models were developed to
describe the probable quark matter proposed at high energy
regimes. In the 1970s, after formulation of QCD, perturbative
calculations of the equations of state of SQM got form, but the
area of validity for these calculations was restricted to very
high densities (Collins \& Perry~\cite{Collins1975}). The
existence of SQSs was also discussed by Witten (1984), who
conjectured that a first order QCD phase transition in the early
universe could concentrate most of the quark excess in dense quark
nuggets. Witten proposed that SQM composed of light quarks is
more stable than nuclei, therefore SQM can be considered as the
ground state of matter.

An SQS would be the bulk SQM phase consisting of almost equal numbers
of up, down and strange quarks, plus a small number of electrons
to ensure charge neutrality. A typical electron fraction is less
than $10^{-3}$ and decreases from the surface to the center of
an SQS (Haensel et al.~\cite{Haensel2007};
Glendenning~\cite{Glendenning2000}; Weber~\cite{Weber1999}; Camenzind~\cite{Camenzind2007}). SQM
would have a lower charge-to-baryon ratio compared to the nuclear
matter and can show itself in the form of an SQS
(Witten~\cite{Witten1984}; Alcock et al.~\cite{Alcock1986};
Haensel et al.~\cite{Haensel1986}; Kettner et
al.~\cite{Kettner1995} ). The collapse of a massive star could
lead to the formation of an SQS. An SQS may also be formed from a
neutron star and is denser than the neutron star (Bhattacharyya et
al.~\cite{Bhattacharyya2006}). If sufficient additional matter is
added to an SQS, it will collapse into a black hole. Neutron stars
with masses of $1.5-1.8 M_{\,\odot}$ with rapid spins are
theoretically the best candidates for conversion to an SQS. An
extrapolation based on this indicates that up to two quark-novae
occur in the observable universe each day. In addition, recent
Chandra observations indicate that objects $RX J185635-3754$ and
$3C58$ may contain SQSs (Prakash et al.~\cite{Prakash2003}). Other
investigations also show that the object $SWIFT J1749.4-2807$ may
be an SQS (Yu \& Xu~\cite{Yu2010}).

The strange quark star, founded from quark matter theory consists
of too many unsolved puzzles which are usually involved in physics of
these relativistic objects. System complexity of these stars
prohibit us from considering all physical and astrophysical
properties simultaneously, and it is possible that some parameters entering the
equation of state do not represent specific physical properties. For example, in MIT
bag model, one of the models used in this paper, when the
researchers try to find and fit the bag constant according to
informations achieved from big colliders (Jin \& Jenning~\cite{Jin1997}; Alford et
al.~\cite{Alford1998};
Blaschke et al.~\cite{Blaschke1999}; Burgio et al.~\cite{Burgio2002b};
Begun et al.~\cite{Begun2011}), we should keep this principle as a
matter of fact that different parameters like
temperature, electromagnetic intensity, density etc. are
important enough on final interpretation for theoretical
calculated bag constant. In this point of view even constant
values of bag pressure no more can be considered purely as the
energy density difference between the perturbative vacuum and the
true vacuum. The role of bag constant for confining quark matter
in comparison with gravity confinement for neutron matter may
require more attention when we consider it for compact stars. Therefore it is better
to consider the dynamical properties of the parameters for investigating of the properties of quarks.
Many works have been done to adapt the theory of
bag Model on physics of ultra-dense matter like using a
density dependent bag constant (Burgio et al.~\cite{Burgio2002a}),
utilizing different values of coupling constants for one gluon
exchange (Farhi \& Jaffe~\cite{Farhi1984}; Berger \&
Jaffe~\cite{Berger1987} ), or instead considering dynamical mass
as effective interaction between particles (Peng et
al.~\cite{Peng1999}; Shao et al.~\cite{Shao2011}).

From perturbative QCD, we know that quarks at ultra high densities
asymptotically interact. One
way of considering the interaction is to assume that quarks exchange
one gluon. Therefor we can add a term to equation of state
that is characterized by a coupling constant. But constant values
of this parameter will weaken the power of interaction in lower
densities while in higher densities, it increases it.
One method to solve the problem is to assume a density dependent
quark mass as the effective interaction. This approach was
investigated in references (Fowler et al.~\cite{Fowler1981};
Chakrabarty et al.~\cite{Chakrabarty1989};
Chakrabarty~\cite{Chakrabarty1991}, \cite{Chakrabarty1994};
Benvenuto \& Lugones~\cite{Benvenuto1995}; Lugones \&
Benvenuto~\cite{Lugones1995}). This has been done by adding a term
to the rest mass which is characterized by a free parameter determined
by stability conditions. They concluded that the density
dependent mass is flavor independent and the applied free
parameter has the same meaning as the bag constant. Then by selecting
one value of bag constant for all densities and flavors, they
tried to obtain the equation of state of quark matter (Peng et
al.~\cite{Peng1999}). A better approach closer to current
work, is to find a solution for density dependent mass from
Nambu-Jona-Lasinio (NJL) method (Carol~\cite{Carol2009}). Carol
calculated the equation of state and structure for hybrid stars
within MIT bag model, while numerical values of density dependent
mass entering in the energy equation had been obtained from
dynamical calculations of mass in NJL model. These numerical
values were entered directly in the pressure equation without
considering density dependency. Quark masses and NJL constants
were also approximate values. The bag constant in that work
was density independent; therefore in addition to previous known
problems of constant values for this parameter (Baldo et
al.~\cite{Baldo2006}; Alford \&
Reddy~\cite{Alford2003}; Alford et al. \cite{Alford2005}), it
misinterprets the meaning of the effective
interaction in some densities.

In Our previous work we considered a hot strange star just after
the collapse of a supernova (Bordbar et
al.~\cite{Bordbar2011vol54}), at finite
temperature with a density dependent bag constant. The
calculations for the structure properties of the strange star at
different temperature indicates that it's maximum mass decreases
by increasing the temperature. In another work (Bordbar \& Peivand~\cite{Bordbar2011vol11}), we
concentrated on the calculation of a bulk of spin polarized SQM at
zero temperature in the presence of a strong magnetic field. We computed
structure properties of this system and found that the presence of
a magnetic field leads to a more stable SQS when compared to
the structure properties of an unpolarized SQS.
In present paper, we investigate the quark matter equation of
state and the strange quark star structure following Carroll
(Carol~\cite{Carol2009}). We base our calculations on MIT bag
model, and after following NJL formalism we extrapolate a density
dependent equation from numerical values of dynamical mass
obtained using NJL method. In Sec. \ref{Energy calculation for
SQM}, the required equations for the MIT bag model are written, the same has been done for NJL model. In Sec.
\ref{MIT bag model with Dynamical mass}, we describe the formalism
applied in this article, and after solving TOV equations in Sec.
\ref{Calculation of strange quark star structure }, we calculates
SQS structure for our method.

\section{Calculation of equation of state for SQM}
\label{Energy calculation for SQM}
In this section, we calculate the equation of state of strange
quark matter (SQM) using MIT and NJL methods as well as MIT method
with the dynamical mass. At first, we introduce these
three models in three separate sections, then we give our
results for the energy and the equation of state of SQM in sec.
\ref{2.4}.
\subsection{ The MIT Bag Model}
\label{The MIT Bag Model} Total energy of a bulk of deconfined up
($u$), down ($d$) and strange ($s$) quarks within MIT bag model is
as follows (Witten~\cite{Witten1984}; Farhi \&
Jaffe~\cite{Farhi1984}; Baym~\cite{Baym1985}; Baym et al.~\cite{Baym1985};
 Berger \& Jaffe~\cite{Berger1987};
Glendenning~\cite{Glendenning1990}; Maruyama et al.~\cite{Maruyama2007}):
\begin{eqnarray} \label{1.1}
\varepsilon = \varepsilon_u+ \varepsilon_d+\varepsilon_s + B.
\end{eqnarray}
In Eq. (\ref{1.1}), $B$ is the bag constant, and
\begin{eqnarray} \label{1}
\varepsilon_{f}\left(\rho_{f}\right)=\frac{3m_{f}\,^{4}}
{8\pi^{2}}\left[x_{f}\left(2x_{f}^{2}
+1\right)\left(\sqrt{1+x_{f}^{2}}\right)-arcsinh\ x_{f}
\right]\nonumber\\
-\alpha_{c}\,\frac{m_{f}\,^{4}}{\pi^{3}}\left[x_{f}^{4}
-\frac{3}{2}\left[x_{f}\left(\sqrt{1
+x_{f}^{2}}\right)-arcsinh\ x_{f}\right]^2\right],
\end{eqnarray}
where $f$ denotes the flavor of the relevant quark, $\alpha_{c}$
is QCD coupling constant and the following term
demonstrates the one-gluon-exchange interaction. In above
equation, $x_{f}$ is defined as follows,
\begin{equation}\label{2}
x_{f}=k_{F}\,^{\left(f\right)}/m_{f},
\end{equation}
where the Fermi momentum $k_{F}\,^{(f)}$  is given by
\begin{equation}\label{3}
k_{F}\,^{(f)}=\left(\rho_{f}\,\pi^{2}\right)^{1/3}
\end{equation}
For the bag constant ($B$), we use a density dependent Gaussian
parametrization (Burgio et al.~\cite{Burgio2002a}; Baldo et
al.~\cite{Baldo2006}):
\begin{equation}\label{4}
B\left(\rho\right)=B_{\infty}+\left(B_{0}-B_{\infty}\right)
\exp[-\beta\left(\rho/\rho_{0}\right)^{2}]
\end{equation}
with $B_{\infty}=B\left(\rho=\infty\right)=8.99\: MeV/fm^{3}, B_{0}=B\left(\rho=0\right)=400\: MeV/fm^{3}$ and
$\beta=0.17$.  In SQM, the beta-equilibrium and charge neutrality
conditions lead to the following relation for the number density
of quarks,
\begin{equation}\label{5}
\rho=\rho_{u}=\rho_{d}=\rho_{s}
\end{equation}
From the total energy, we can obtain the equation of state of SQM
using the following relation,
\begin{equation}\label{6}
P(\rho)=\rho\frac{\partial\varepsilon}{\partial\rho}-\varepsilon.
\end{equation}
\subsection{The Nambu-Jona-Lasinio Model}
\label{The Nambu-Jona-Lasinio Model} Here we give a brief
introduction regarding the calculations in the Nambu-Jona-Lasinio
(NJL) method. For NJL model, we use a common three flavor
lagrangian adopted from (Rehberg et al.~\cite{Rehberg1996})
which preserves chiral symmetry of QCD,
\begin{eqnarray} \label{7}
{\mathcal
L}=\bar{q}\left(i\gamma^{\mu}\partial_{\mu}-\hat{m_{0}}\right)q+G{\textstyle
{\displaystyle
\sum_{k=0}^{8}\left[\left(\bar{q}\lambda_{k}q\right)^{2}
+\left(\bar{q}i\gamma_{5}\lambda_{k}q\right)^{2}\right]-}}
\\ \nonumber K\left[det_{f}\left(\bar{q}\left(1+\gamma_{5}\right)q\right)
+det_{f}\left(\bar{q}\left(1-\gamma_{5}\right)q\right)\right].
\end{eqnarray}
In adopted lagrangian, $q$ denotes quark field with three flavors
$u$, $d$ and $s$, and three colors. $\hat{m_{0}}=diag(m_{0}^{u},\,
m_{0}^{d},\, m_{0}^{s})$ is a $3\times3$ matrix in flavor space.
And $\lambda_{k}$ ( $0\leq k \leq 8$ ) are the $U(3)$ flavor matrices.
We restrict ourselves to the isospinsymmetric case,
$m_{0}^{u}=m_{0}^{d}$. We have picked up the parameters from
references (Kunihiro~\cite{Kunihiro1989}; Ruivo et al.~\cite{Ruivo1999}; Buballa \& Oertel~\cite{Buballa1999})
 which are fitted to the pion mass,
the pion decay constant, the kaon mass and the quark condensates.

\begin{figure}

\includegraphics[width=8.1cm]{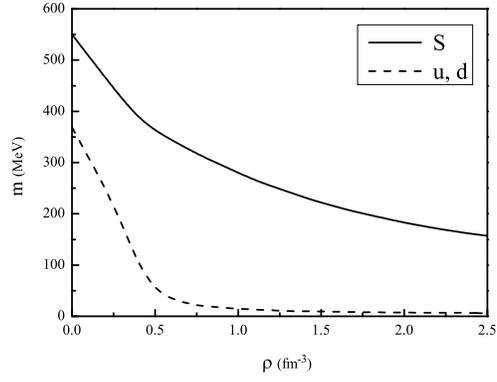}

\caption{Density dependent mass ($m$) versus density ($\rho$)
obtained from dynamical NJL model.}

\label{quarkmass}
\end{figure}

NJL model is an unrenormalizable method with divergent
integrations. To
prevent the divergence, we need to introduce some breaking points
for upper limit of integrals which satisfy the physics ranges of
our problem. It is usually done by choosing a proper cut-off. In
present paper, the adopted cut-off is named Ultra-violet cut-off
that indicates restoring of chiral symmetry breaking,
$\Lambda=602.3\ MeV$. $G$ and $K$ are
coupling strengths that read,  $G\Lambda^{2}=1.835,\:
K\Lambda^{5}=12.36$. The rest mass of $s$ quark is
$m_{0}^{s}=140.7\ MeV$, and there is $m_{0}^{u}=m_{0}^{d}=5.5\
MeV$ for $u$ and $d$ quarks. The baryon number density is given by
\begin{equation}\label{8}
\rho_{B}=\frac{1}{3}n_{B}=\frac{1}{3}\left(n_{u}+n_{d}+n_{s}\right),
\end{equation}
where $n_{i}=\left\langle q{}_{i}^{\dagger}q_{i}\right\rangle$.
Within mean field approximation, the dynamical mass is calculated
by the following gap equation,
\begin{equation}\label{9}
m_{i}=m_{0}^{i}-4G\left\langle \bar{q_{i}}q_{i}\right\rangle
+2K\left\langle \bar{q_{j}}q_{j}\right\rangle \left\langle
\bar{q_{k}}q_{k}\right\rangle.
\end{equation}
In the above equation, we need to calculate permutation of all quark
flavors. The quark condensate in Eq. (\ref{9}) reads
\begin{equation}\label{10}
\left\langle \bar{q_{i}}q_{i}\right\rangle
=-\frac{3}{\pi^{2}}{\int}_{P_{Fi}}^{\Lambda}P^{2}dp\frac{m_{i}}{\sqrt{m_{i}^{2}+p^{2}}},
\end{equation}
and $P_{Fi}$, Fermi momentum of quark $i$, is obtained from the
following relation,
\begin{equation}\label{11}
P{}_{Fi}=\left(\pi^{2}n_{i}\right)^{\frac{1}{3}}.
\end{equation}
Equations (\ref{9}) and (\ref{10}) have self consistent solutions.
It means that for a given number density, $n_{i}$, we should
calculate quark condensate and substituting the corresponding value in Eq. (\ref{9})
to reach a consistent result of the dynamical mass after doing the iteration process.
In Fig. \ref{quarkmass}, we have
plotted the results of density dependent mass for $u$, $d$ and $s$
quarks as a function of density. As it is clear from Fig.
\ref{quarkmass}, quark masses vary from current masses ($5.5MeV$
for $u$ and $d$ quarks, and $140.7MeV$ for $s$ quark) at high
densities to constituent mass at near zero densities ($368.7MeV$
for $u$ and $d$ quarks, and $550MeV$ for $s$ quark).

\begin{figure}

\includegraphics[width=8.1cm]{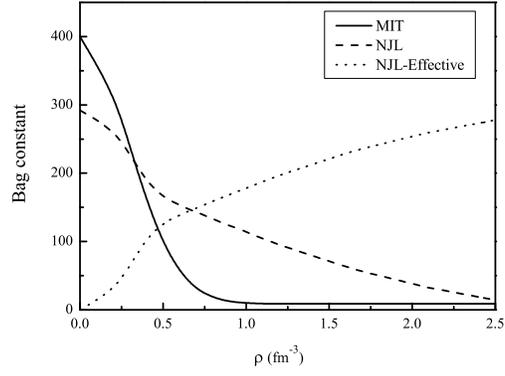}
\caption{Bag constant as a function of density for NJL and MIT
models.} \label{bag}
\end{figure}

The solution via mean field approximation forces us to stabilize
equations by diminishing energy density and pressure in vacuum.
This is satisfied by entering a parameter which has the same
meaning of bag constant in MIT bag model (Buballa \&
Oertel~\cite{Buballa1999}):
\begin{eqnarray}\label{12}
B=\sum_{i=u,d,s}\left(\frac{3}{\pi^{2}}{\int}_{0}^{\Lambda}\;
p^{2}dp\left(\sqrt{p^{2}+m_{i}^{2}}-\sqrt{p^{2}
+{m^{i}}_{0}^{2}}\right)
-2{G\left\langle \bar{q_{i}}q_{i}\right\rangle}^{2}\right)\nonumber \\
+ \;4K\left\langle \bar{u}u\right\rangle\left\langle
\bar{d}d\right\rangle\left\langle \bar{s}s\right\rangle
\end{eqnarray}
Now we can calculate the equation of state of SQM in NJL model,
\begin{equation}\label{13}
p=-\varepsilon+\sum_{i=u,d,s}n_{i}\sqrt{P_{F}{}_{i}^{2}+m_{i}^{2}},
\end{equation}
where
\begin{equation}\label{13}
\varepsilon=\sum_{i=u,d,s}\frac{3}{\pi^{2}}
{\int}_{0}^{P_{Fi}}p^{2}dp\sqrt{p^{2}+m_{i}}-(B-B_0).
\end{equation}
Parameter $B$ is the bag pressure, which is explained by Buballa
(2005), and is a dynamical consequence of the mean field solution, not
a parameter inserted by hand, as was done in MIT bag model.
It is shown in Fig. \ref{quarkmass}, matter in
NJL method acquires dynamical mass in nonzero baryon densities,
but in MIT bag model, the given mass remains constant for all
densities. Consequently, this will lead to dissimilar chiral
symmetry behavior as density changes. In NJL model, since quarks
acquire dynamical mass, the chiral symmetry spontaneously breaks
in lower densities, while in MIT bag model, it will happen
physically when quarks change their directions by hitting the bag
(what is not considered theoretically in ordinary MIT bag model).
The bag constant versus density is presented in Fig. \ref{bag} for
our used models. It is apparent from Fig. \ref{bag} that chiral
symmetry in our calculations is fully restored in the densities
greater than $\rho\simeq2.5\ fm^{-3}$.
It is also important to mention that vacuum in MIT bag model is
totally free of particles (flow of particle's wave function is
restricted by the confinement), while in NJL model no confinement
is produced. In other word, the vacuum in NJL model is made of
paired quasi-quarks that lower the energy density of particles in
comparison to MIT bag model. From the above discussions, it seems reasonable
to add an effective bag constant to energy equation
(Buballa~\cite{Buballa2005}),
\begin{eqnarray}\label{14}
B_{0}=B\mid_{n_{u}=n_{d}=n_{s}=0}, \nonumber\\ B_{eff}=B-B_{0}.
\end{eqnarray}
From Fig. \ref{bag}, it seems that the effective bag constant
diminishes at zero density. Then the correct interpretation for
the effective bag constant is the energy per volume needed to
fully break quark-antiquark pairs in order to completely restore
chiral symmetry at ultra high densities. Even the maximum value of
dynamical NJL bag constant is smaller than that of MIT's one, because it
reduces the energy per particle due to quark-antiquark pairing at
lower densities (Buballa~\cite{Buballa2005}). Fig. \ref{bag} shows
that the decreasing rate of
MIT bag constant is higher than that of NJL. This indicates that
MIT bag model does gross approximation over physics of matter in
middle and higher densities $(\rho>0.8\ fm^{-3})$. Therefore, the
density dependent bag constant should be corrected by another
higher density sensitive parameter. This could not be achieved
by a one gluon exchange term that considers the interaction with a
constant strength in all energy regimes.
Fig. \ref{bag} indicates that at the density $\rho\simeq 0.45\
fm^{-3}$, there is a cross point for the effective bag constant of NJL model and
the bag constant of MIT model.
As it is mentioned in above discussions, the bag pressure is
the energy needed to confine particles where effective bag constant is
energy needed to destabilize quark-antiquark pairs. Now, we can
suggest that the hadron-quark phase transition can takes place at
the density $\rho\simeq0.45\ fm^{-3}$. This is in good agreement
with the results of others (Heinz~\cite{Heinz2001}; Heinz \& Jacob~\cite{Heinz2000}).

\subsection{  MIT bag model with dynamical mass   }
\label{MIT bag model with Dynamical mass}

In MIT bag model with dynamical mass, we consider the effect
of dynamical behavior of the quark mass in calculating the equation of
state of SQM within MIT bag model using NJL numerical mass
results. In fact, we use the dynamical masses (Fig.
\ref{quarkmass}) for $u$, $d$ and $s$ quarks in Eq. (\ref{1})
instead of their fixed values.

\subsection{Our results for the energy and equation of state of SQM}
\label{2.4}

To distinguish numerous outcomes, we present the results of our calculations in three
following models;
\begin{itemize}
\item Model 1: MIT model by a density dependent bag constant
and one gluon-exchange $(\alpha_{c}=0,\,0.16,\,0.5)$ as effective
interaction.

\item Model 2: NJL model.

\item Model 3: MIT bag model by a density dependent bag
constant, dynamical mass and one gluon-exchange $(\alpha_{c}=0,\,0.16,\,0.5)$ as effective interaction.
\end{itemize}

\begin{figure}

\includegraphics[width=8.1cm]{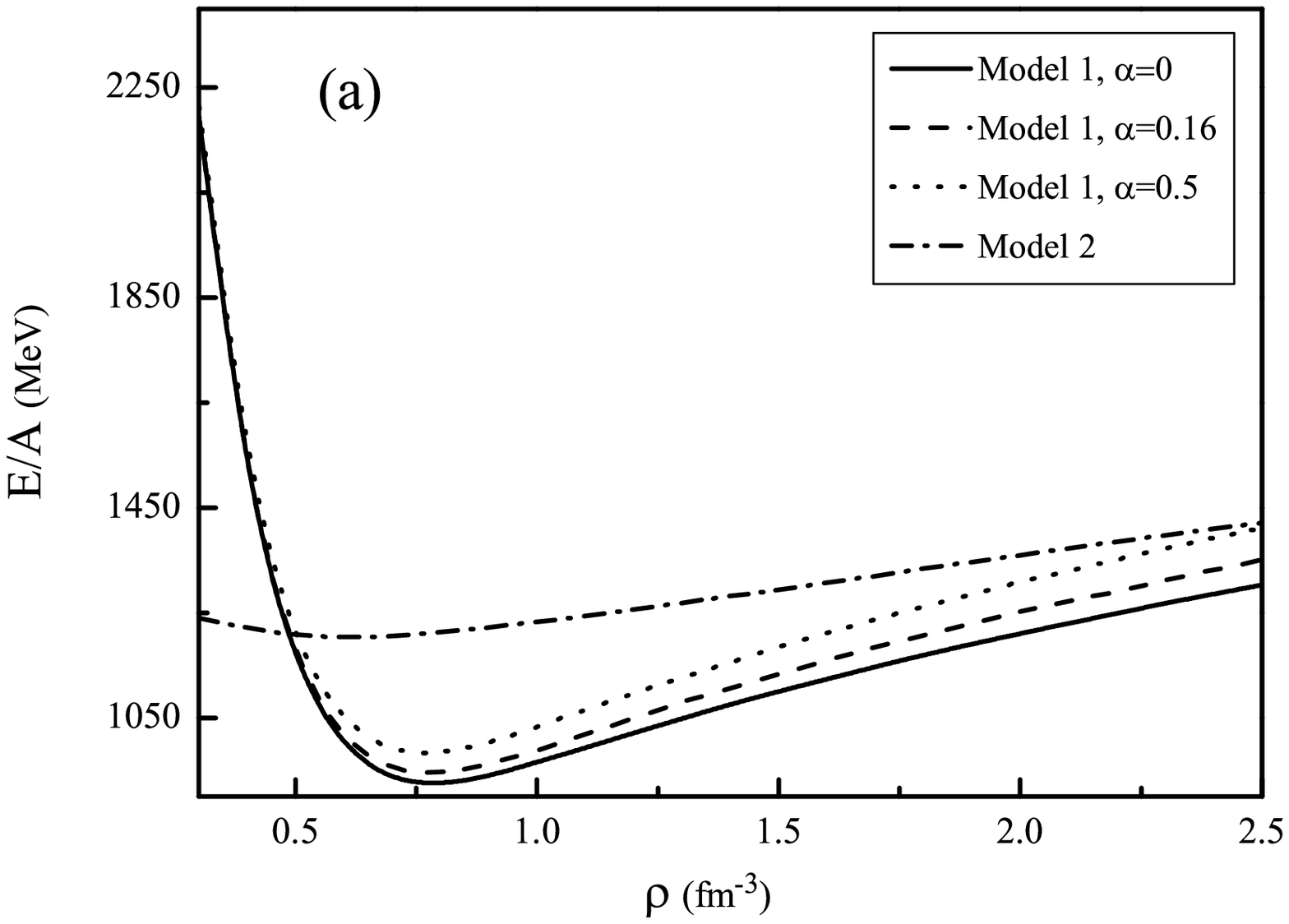}
\includegraphics[width=8.1cm]{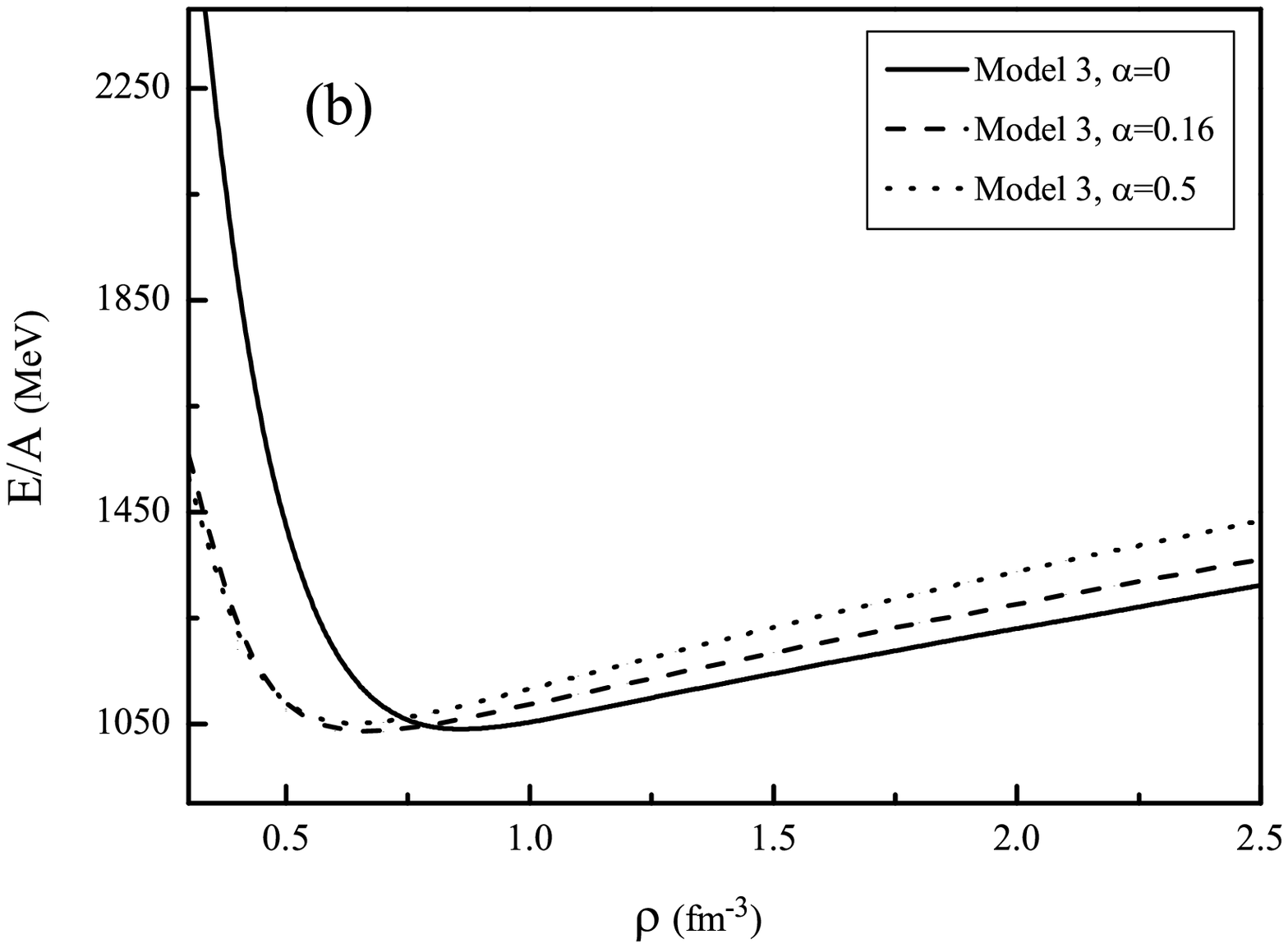}
\caption{The energy per baryon versus density for models 1 and 2 (a), and model 3 (b).} \label{energy}
\end{figure}

Our results for the energy of SQM versus density calculated with
above models have been plotted in Fig. \ref{energy}.
We see that for both MIT based calculations (models 1 and 3), at
lower densities $(\rho<0.5\ fm^{-3})$, the energy of SQM suddenly
increases as the density decreases. This shows the concept of
confinement (Buballa~\cite{Buballa2005}). For these two models, we also see
that the energy of SQM gets to a minimum, then increases with a
small rate.
Fig. \ref{energy} shows that for model 1 and model 3, the energies of
different coupling constants are nearly identical for densities
$\rho<0.5fm^{-3}$. However, they have a substantial difference as
the density increases.
We can see that at lower densities $(\rho<0.7\ fm^{-3})$, the
results of model 3 is considerably different from those of model
1. While this difference becomes small as density increases,
specially for lower values of coupling constant, due to
asymptotic freedoms which is the simple MIT bag model without
interaction.
From Fig. \ref{energy}, it is seen that the energy of SQM in model
2 (NJL model) has finite values even at low densities showing no
confinement.
We also see that the energy of SQM from model 3 with smaller values of coupling constant is lower than
that of model 2 for $\rho>0.7\ fm^{-3}$ indicating a more stable
state of quark matter at these densities. However, at very high
densities, the difference between the results of these two models
becomes negligible.

\begin{figure}
\includegraphics[width=8.1cm]{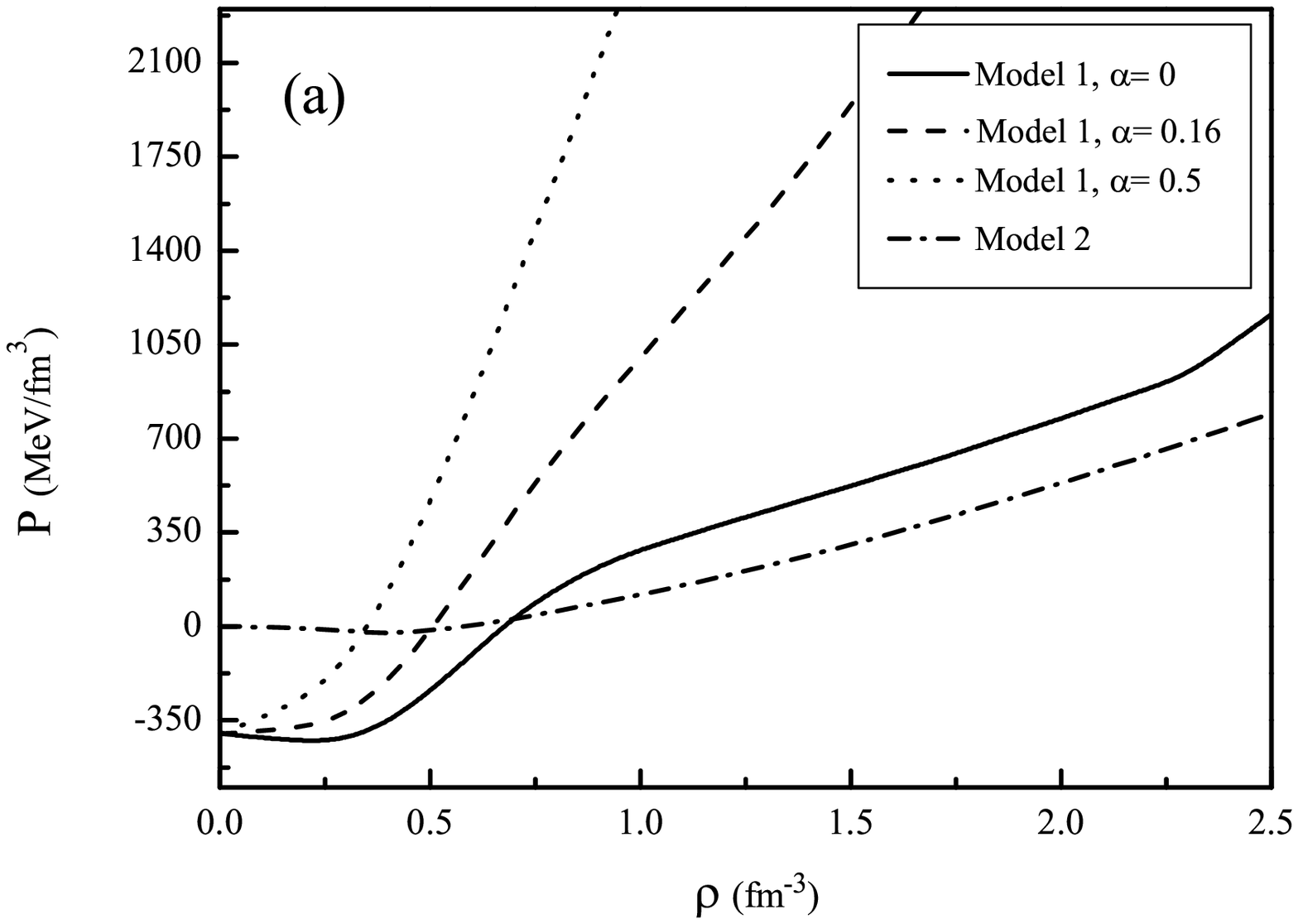}
\includegraphics[width=8.1cm]{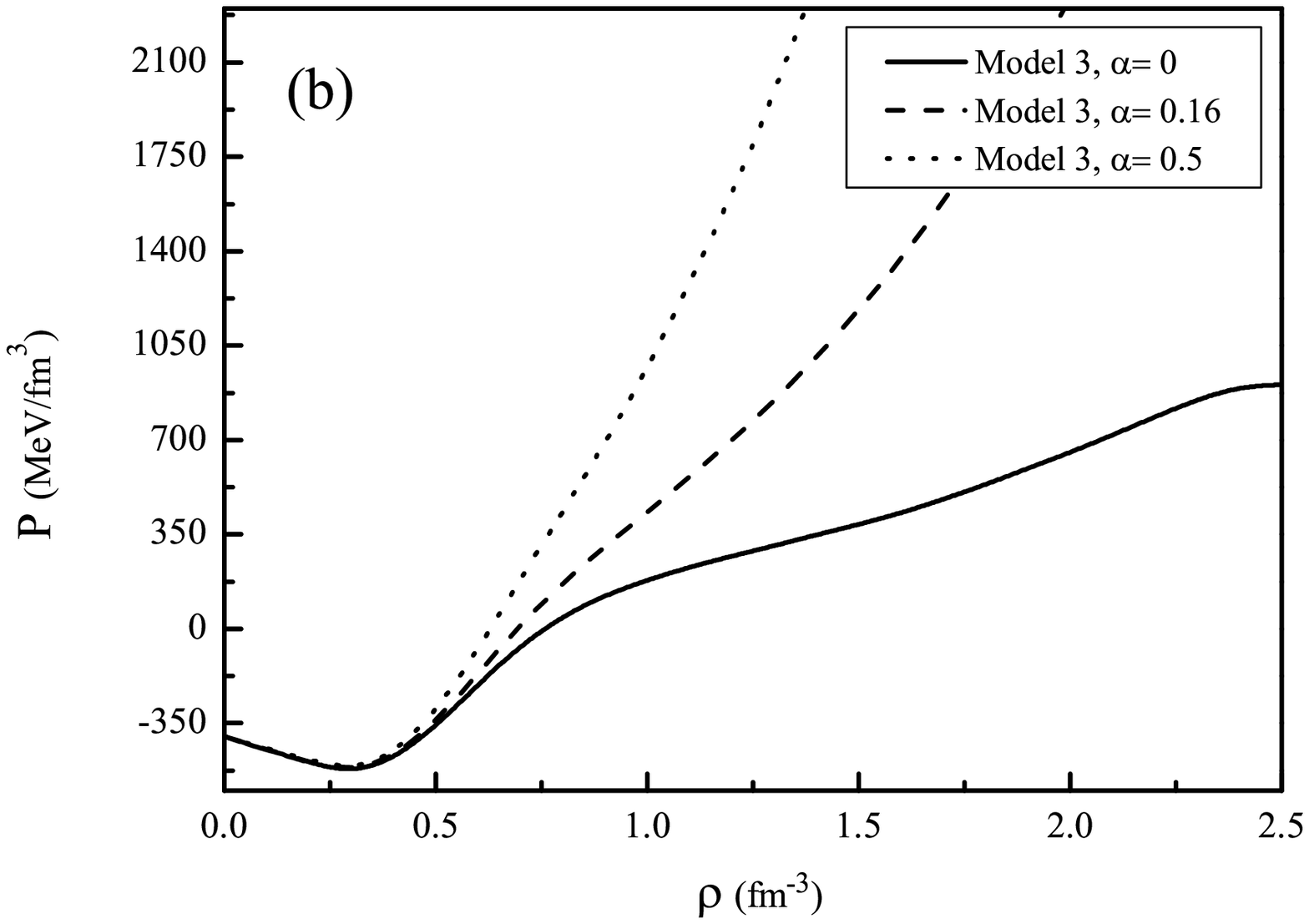}

\caption{Pressure as a function of density for models 1 and 2 (a), and model 3 (b).} \label{pressure}
\end{figure}

In Fig. \ref{pressure}, our results for the pressure of SQM have
been plotted versus density.
It can be found that for MIT bag model, the higher values of
coupling constant leads to the stiffer equation of state for SQM.
Fig. \ref{pressure} shows that by considering a dynamical mass for
the quarks (density dependent mass) in MIT model, we get the lower
values for the pressure of SQM.
For $\alpha_{c}=0.0$, we see that the result of model 3 for the equation of
state of SQM is nearly identical with that of model 1.
It can be seen that for $\rho>0.6\ fm^{-3}$, our results for the
pressure of SQM calculated by NJL model are nearly identical with
those of model 3 and model 1 for $\alpha_{c}=0.0$, while at lower
densities, there is a considerable difference between them.

\begin{figure}

\includegraphics[width=8.1cm]{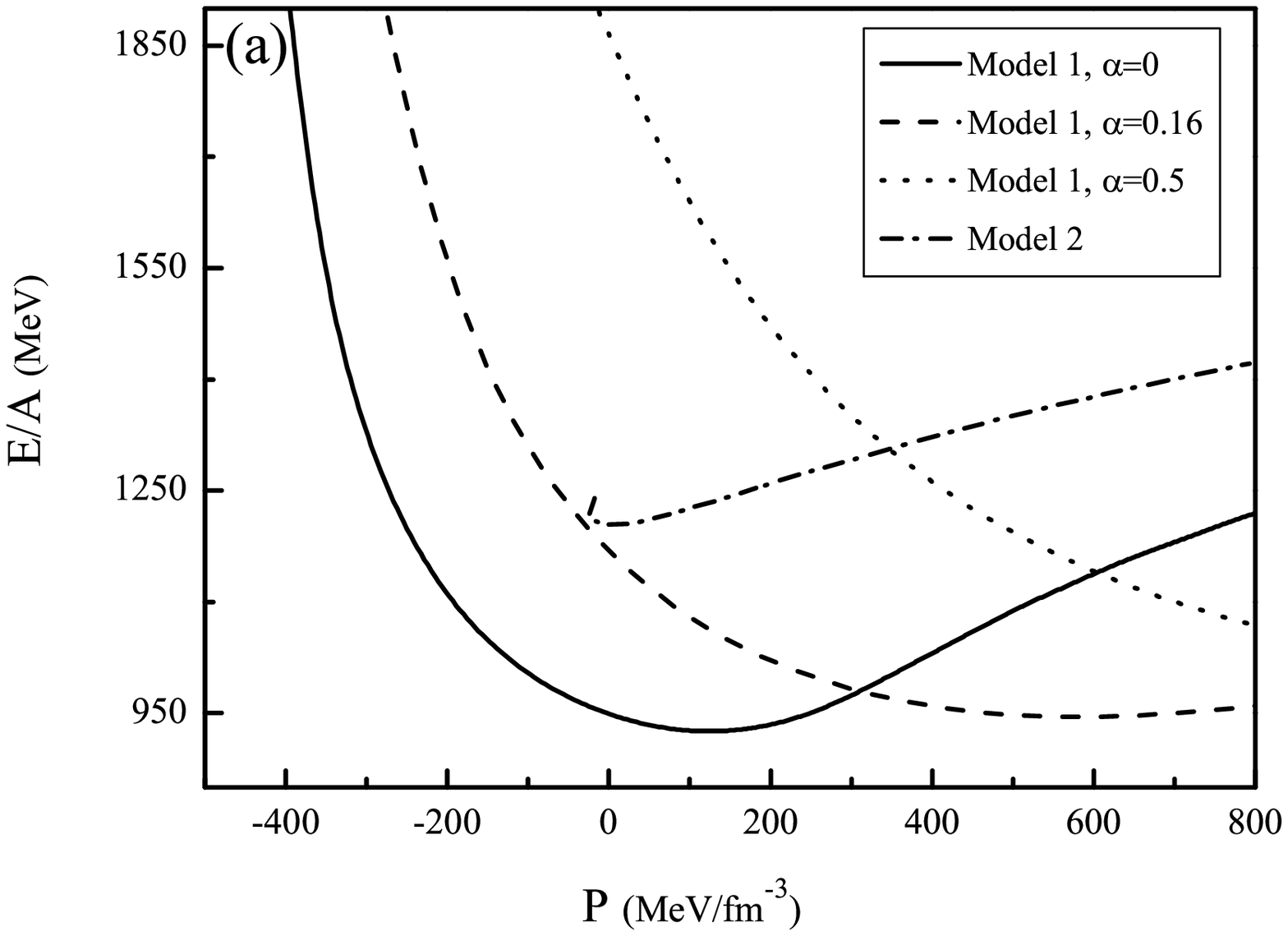}
\includegraphics[width=8.1cm]{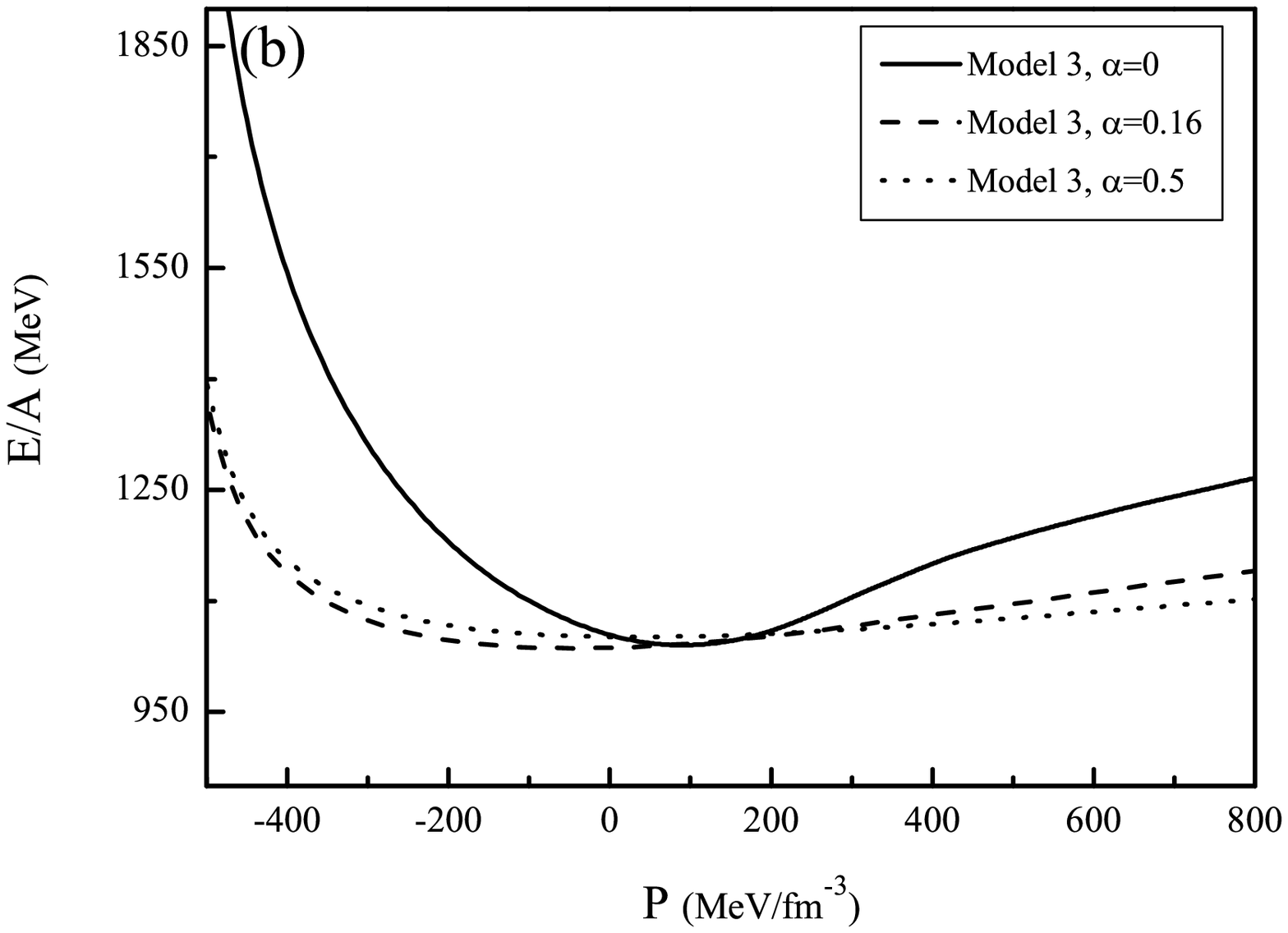}

\caption{Energy per particle versus pressure for models 1 and 2 (a), and model 3 (b).
} \label{e-p}
\end{figure}

In order to investigate the quark matter stability, the energy of
SQM versus pressure has been plotted in Fig. \ref{e-p}.
It is clearly seen that at zero pressure, the MIT bag model with
$\alpha_{c}=0$ leads to the lowest value for the energy of SQM
($950MeV\ fm^{-3}$) compared to other models.  This value is
comparable with the result for the binding energy per particle of
$^{56}Fe$ ($930MeV\ fm^{-3}$) (Witten~\cite{Witten1984}). This indicates that
among different models used in this work, MIT model with
$\alpha_{c}=0$ shows the most stable state of SQM.

\section{  Calculation of strange quark star structure }
\label{Calculation of strange quark star structure } The
gravitational mass ($M$) and radius ($R$) of compact stars are of
special interest in astrophysics. In this section, we calculate
the structural properties of a strange quark star for our three
models. Using the equation of state of strange quark matter for
the models applied in this work, we can obtain $M$ and $R$ by
numerically integrating the general relativistic equations of
hydrostatic equilibrium, the Tolman-Oppenheimer-Volkoff (TOV)
equations, which are as follows (Shapiro \&
Teukolsky~\cite{Shapiro1983}),

\begin{equation}\label{15}
\frac{dm}{dr}=4\pi r^{2}\varepsilon\left(r\right),
\end{equation}

\begin{equation}\label{16}
\frac{dp}{dr}=-\frac{Gm\left(r\right)\varepsilon\left(r\right)}{r^{2}}\left(1
+\frac{p\left(r\right)}{\varepsilon\left(r\right)c^{2}}\right)\left(1
+\frac{4\pi
r^{3}p\left(r\right)}{m\left(r\right)c^{2}}\right)\left(1
-\frac{2Gm\left(r\right)}{c^{2}r}\right)^{-1},
\end{equation}
where $\varepsilon\left(r\right)$ is the energy density, $G$ is
the gravitational constant, and
\begin{equation}\label{16}
m(r)={\int}_{0}^{r}4\pi \acute{r}^2\,\varepsilon(\acute{r})d\acute{r}
\end{equation}
has the interpretation of the mass inside radius $r$. By selecting
a central energy density $\varepsilon_c$, under the boundary
conditions $P(0)=P_c$ and m(0)=0, we integrate the TOV equation
outwards to a radius $r=R$, at which $P$ vanishes.

\begin{figure}

\includegraphics[width=8.1cm]{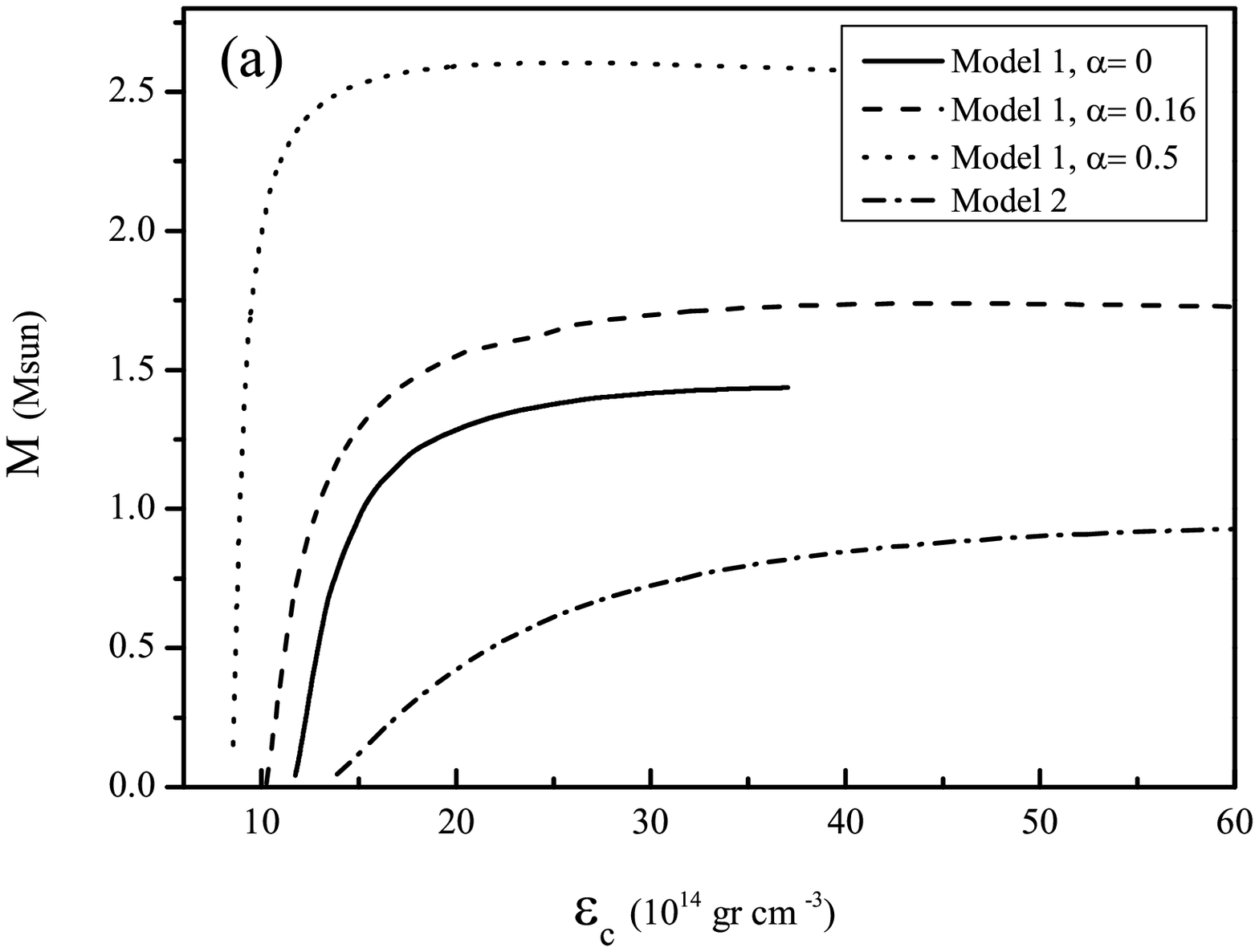}
\includegraphics[width=8.1cm]{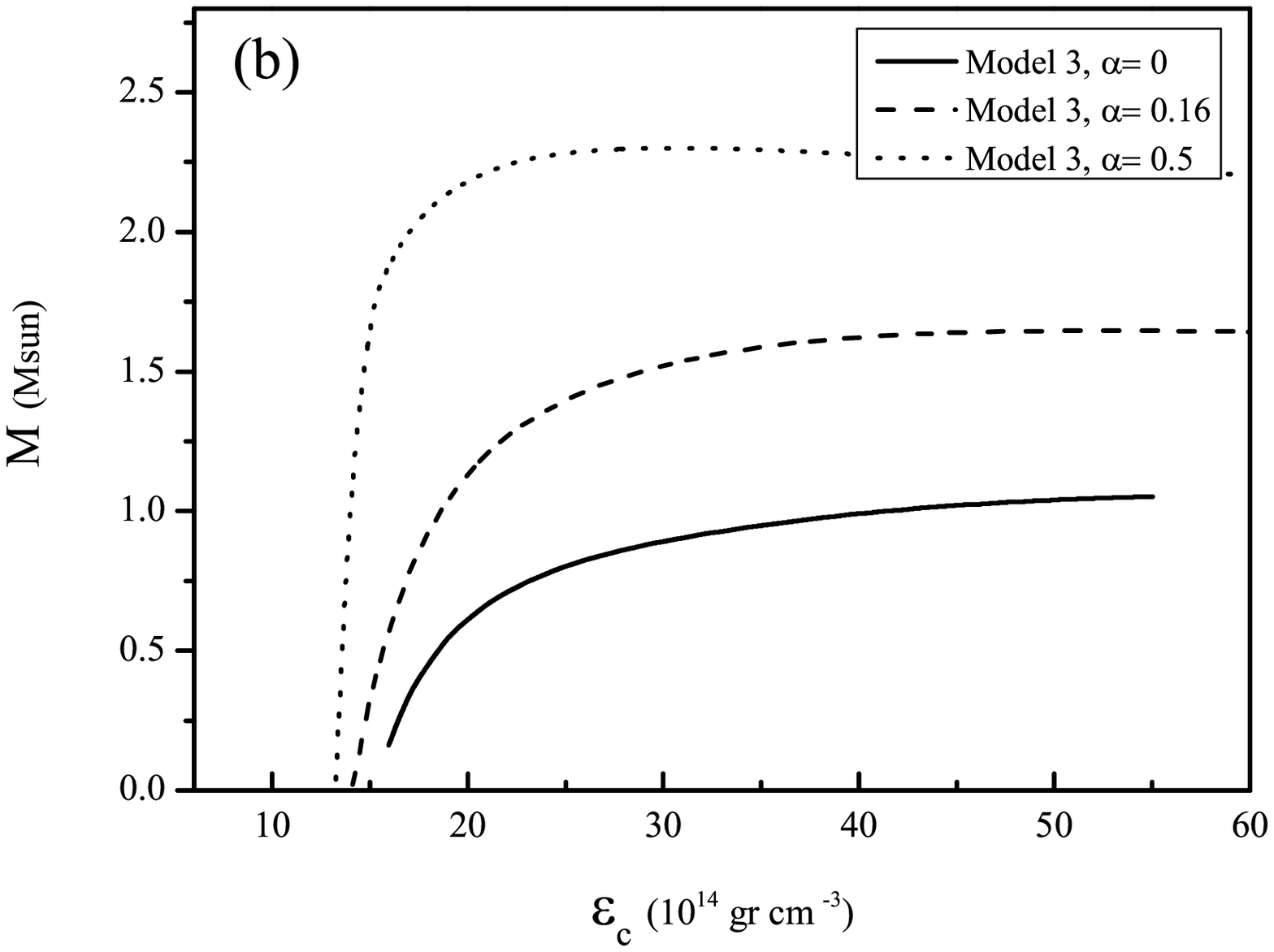}
\caption{Gravitational mass $(M)$ in unit of solar mass $(M_{sun})$ versus
central energy density $(\varepsilon_c)$ for models 1 and 2 (a), and model 3 (b).} \label{mass-energy}
\end{figure}

In Fig. \ref{mass-energy}, we have presented our results for the
gravitational mass of SQS versus the central energy density. Fig.
\ref{mass-energy} shows that at low energy densities, the
gravitational mass increases rapidly by increasing the energy
density, and it finally reaches to a limiting value (maximum
gravitational mass) at higher energy densities. It is seen that
the increasing rate of mass for Model 3 with higher values of coupling constant is substantially higher
than those of other models.
Table \ref{Maximum gravitational mass} summarizes maximum
gravitational masses of different applied models and the
corresponding radii. As it seen from Table \ref{Maximum
gravitational mass}, we can conclude that using dynamical mass in
energy equation and equation of state of SQM reduces the
calculated maximum mass. This is in a good agreement with many observational data
obtained from low mass compact stars (Zhang~\cite{Zhang2007}).
It is interesting that in spite of considering dynamical mass as
the effective interaction in MIT bag model (model 3 with $\alpha_{c}=0$), we find the
smaller SQS maximum mass in comparison to MIT bag model (model 1)
even without interaction ($\alpha_{c}=0$).
%
As it is obvious from Table \ref{Maximum gravitational mass}, for
models 1 and 3, the calculated maximum mass increases as strong coupling
constant increases. This behavior demonstrates that ultra massive
SQS with masses greater than $M=1.05M_{\odot}$ are stars which are
composed of highly interacting strange quark matter.
%
We note that some studies indicate that there exist a big
uncertainty about mass and radius of ultra massive stars with
$M>1.9M_{\odot}$ (Lattimer \& Prakash~\cite{Lattimer2010}). These
studies showed that the observed data of mass and radius for these
stars, which commonly belong to X-ray stars, were wrongly
calculated and the calculations were revised to the smaller values for
mass and radius. The best example is $PSR\: J\,0751+1807$ pulsar
that initially was supposed to have a mass of
$M=2.2\pm0.2M_{\odot}$ but recently revised to $M=1.26M_{\odot}$
(Lattimer \& Prakash~\cite{Lattimer2010}).

\begin{figure}

\includegraphics[width=8.1cm]{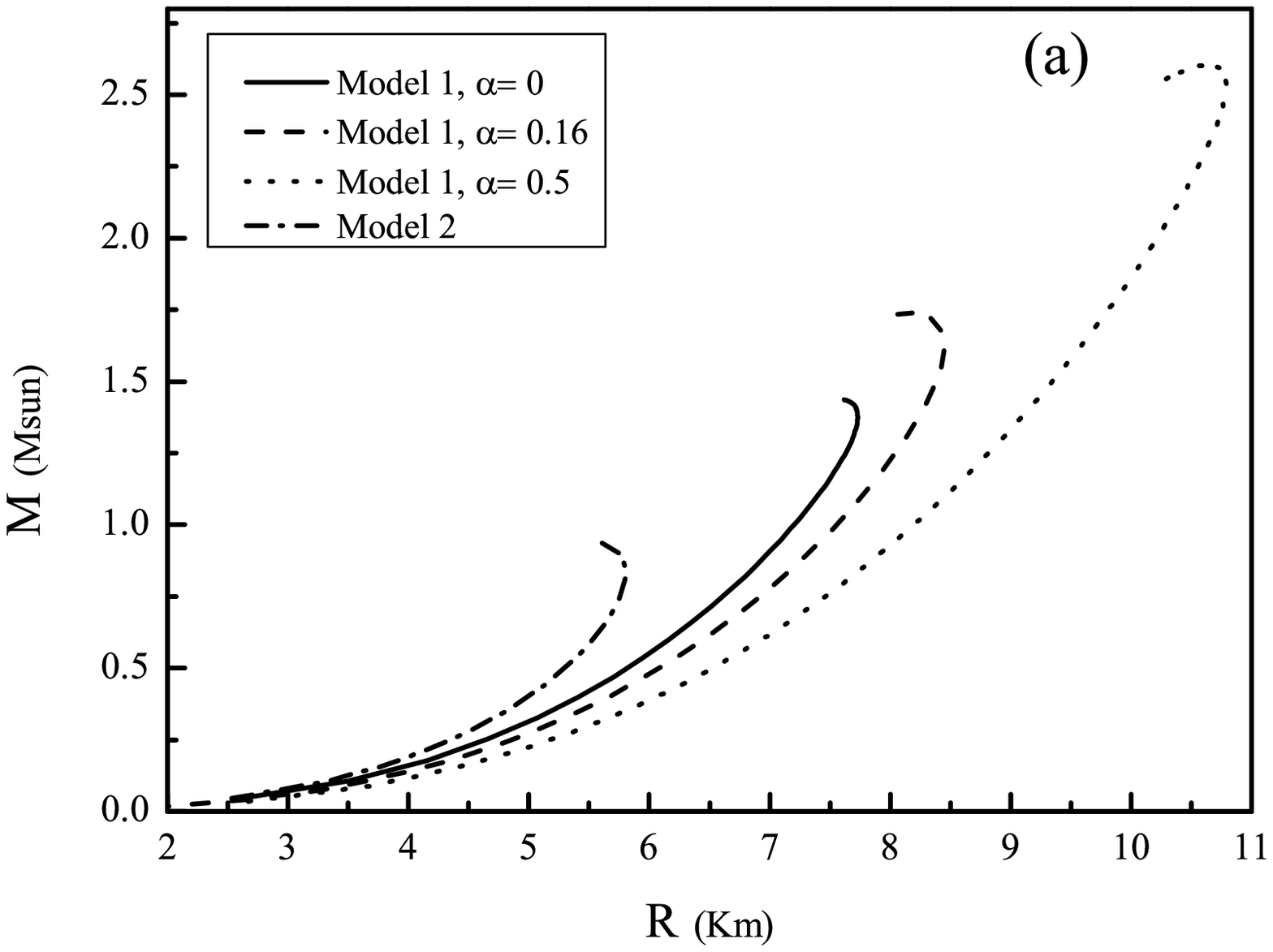}
\includegraphics[width=8.1cm]{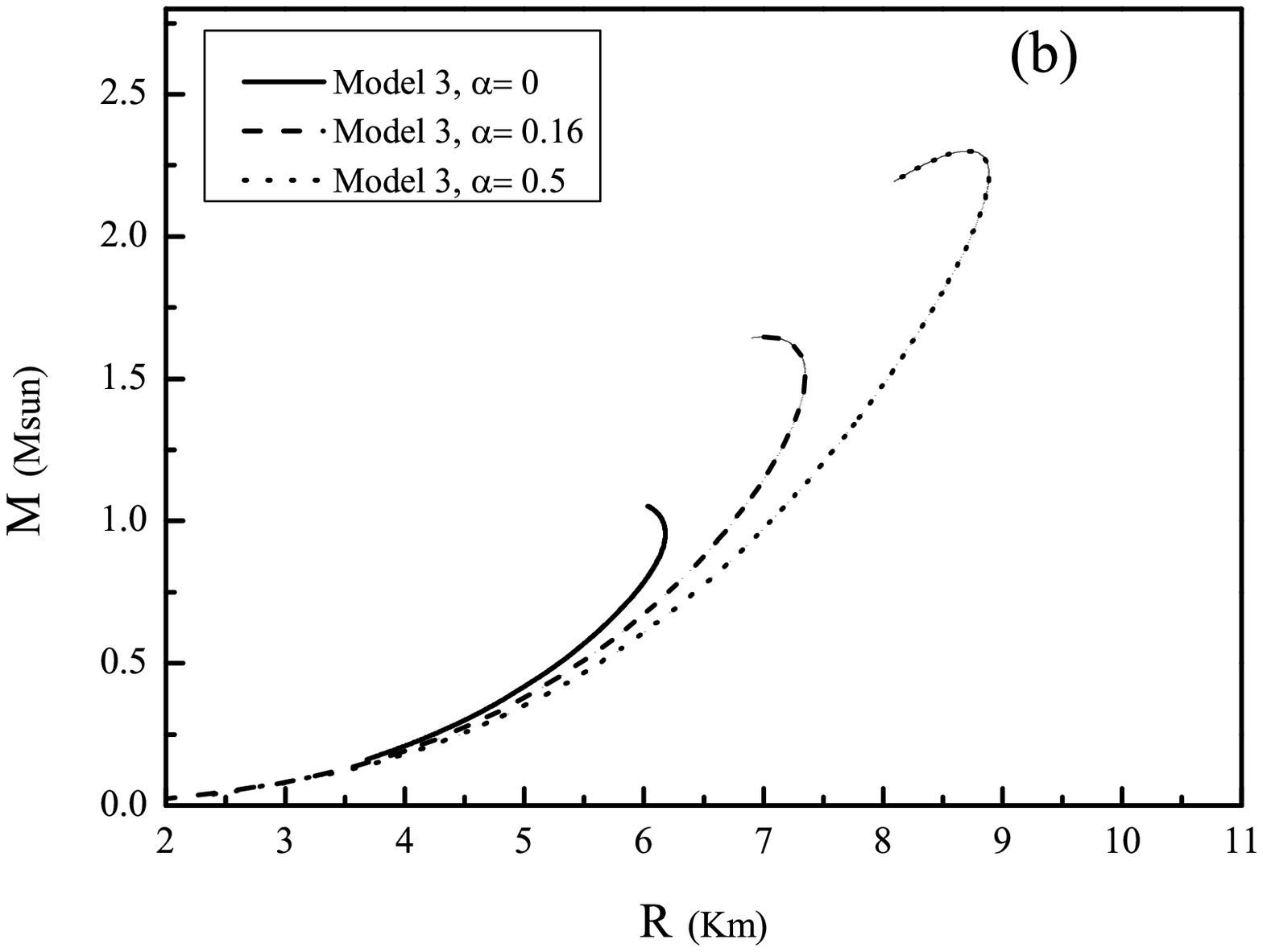}
\caption{Gravitational mass $(M)$ in unit of solar mass $(M_{sun})$ versus
radius $(R)$ for models 1 and 2 (a), and model 3 (b).} \label{mass}
\end{figure}

We have also plotted the gravitational mass of SQS versus radius
for our three models in Fig. \ref{mass}. It is seen that for all
models, the mass increases by increasing the radius, but with
different increasing rates for different models. Fig. \ref{mass}
shows that for a given value of radius, the dynamical model (model
3) gives the smaller mass with respect to that of MIT bag model
(model 1); however, for $\alpha_{c}=0$, it is close to the result of NJL model (model
2).

\section*{Acknowledgements}
{This work has been supported by Research Institute for Astronomy
and Astrophysics of Maragha. We wish to thank Shiraz University
Research Council.}



\newpage
\begin{table}[h]
\begin{center}
  \caption[]{Maximum gravitational mass $(M_{max})$ and the corresponding
  radius $(R)$ for different applied models.}

  \begin{tabular}{clcl}
  \hline\noalign{\smallskip}
 & $M_{max}\left(M_{\,\odot}\right)$ & $R\left(km\right)$  \\
 \hline\noalign{\smallskip}

 Model 1; $\alpha_{c}=0$ & 1.43 & 7.61  \\
 Model 1; $\alpha_{c}=0.16$ & 1.73 & 8.17  \\
 Model 1; $\alpha_{c}=0.5$  & 2.6 & 10.6\\
 Model 2  & 0.98 & 5.59  \\
 Model 3; $\alpha_{c}=0$& 1.05 & 6.03  \\
 Model 3; $\alpha_{c}=0.16$ & 1.65 & 6.98  \\
 Model 3; $\alpha_{c}=0.5$ & 2.3 & 8.69  \\
 \noalign{\smallskip}\hline
  \end{tabular}
\end{center}
\label{Maximum gravitational mass}
\end{table}

\newpage

\newpage
\newpage
\newpage

\newpage
\newpage
\newpage

\end{document}